\documentclass[review]{elsarticle}
\usepackage{color}
\usepackage{amsmath}
\usepackage{amsfonts}
\usepackage{subcaption}
\usepackage{graphicx}

\journal{...}

\newcommand{\pvalue}{{\em p-}value }
\newcommand{\ie}{i.e. }

\newcommand{\lnom}{$-13.37\;$}  
\newcommand{\lshift}{$-17.05$} 
\newcommand{\dlmeas}{3.68 }

\newcommand{\AMnom}{$19$} 
\newcommand{\AMnomerr}{$\pm5$} 
\newcommand{\AMshift}{$20$} 
\newcommand{\AMshifterr}{$\pm5$} 

\newcommand{\pvHzero}{$2.6\times10^{-4}\;$} 
\newcommand{\soneHzero}{$3.5\,\sigma$}

\newcommand{\pvdelta}{$0.23\;$} 
\newcommand{\stwodelta}{$1.2\,\sigma\;$} 

\begin{document}

\begin{frontmatter}

\title{The cosmic ray shadow of the Moon observed with the ANTARES neutrino telescope}

\author[IPHC]{A.~Albert}
\author[UPC]{M.~Andr\'e}
\author[Genova]{M.~Anghinolfi}
\author[Erlangen]{G.~Anton}
\author[UPV]{M.~Ardid}
\author[CPPM]{J.-J.~Aubert}
\author[APC]{J.~Aublin}
\author[APC]{T.~Avgitas}
\author[APC]{B.~Baret}
\author[IFIC]{J.~Barrios-Mart\'{\i}}
% ...

\address[IPHC]{\scriptsize{Universit\'e de Strasbourg, CNRS,  IPHC UMR 7178, F-67000 Strasbourg, France}}
\address[UPC]{\scriptsize{Technical University of Catalonia, Laboratory of Applied Bioacoustics, Rambla Exposici\'o, 08800 Vilanova i la Geltr\'u, Barcelona, Spain}}
\address[Genova]{\scriptsize{INFN - Sezione di Genova, Via Dodecaneso 33, 16146 Genova, Italy}}
\address[Erlangen]{\scriptsize{Friedrich-Alexander-Universit\"at Erlangen-N\"urnberg, Erlangen Centre for Astroparticle Physics, Erwin-Rommel-Str. 1, 91058 Erlangen, Germany}}
\address[UPV]{\scriptsize{Institut d'Investigaci\'o per a la Gesti\'o Integrada de les Zones Costaneres (IGIC) - Universitat Polit\`ecnica de Val\`encia. C/  Paranimf 1, 46730 Gandia, Spain}}
\address[CPPM]{\scriptsize{Aix Marseille Univ, CNRS/IN2P3, CPPM, Marseille, France}}
\address[APC]{\scriptsize{APC, Univ Paris Diderot, CNRS/IN2P3, CEA/Irfu, Obs de Paris, Sorbonne Paris Cit\'e, France}}
\address[IFIC]{\scriptsize{IFIC - Instituto de F\'isica Corpuscular (CSIC - Universitat de Val\`encia) c/ Catedr\'atico Jos\'e Beltr\'an, 2 E-46980 Paterna, Valencia, Spain}}
% ...
\address[Genova-UNI]{\scriptsize{Dipartimento di Fisica dell'Universit\`a, Via Dodecaneso 33, 16146 Genova, Italy}}
% ...


\begin{abstract}

One of the main objectives of  the ANTARES telescope is the search for point-like neutrino sources. Both the pointing accuracy and the angular resolution of the detector are important in this context and a reliable way to evaluate this performance is needed.
In order to measure the pointing accuracy of the detector, one possibility is to study the shadow of the Moon, \ie the deficit of the atmospheric muon flux from the direction of the Moon induced by the absorption of cosmic rays.
Analysing the data taken between 2007 and 2016, the Moon shadow is observed with $3.5\sigma$ statistical significance. The detector angular resolution for downward-going muons is 0.73$^{\circ}\pm0.14^{\circ}.$ The resulting pointing performance  is consistent with the expectations.
An independent check of the telescope pointing accuracy is realised with the data collected by a shower array  detector onboard of a ship temporarily moving around the ANTARES location.

\end{abstract}

\end{frontmatter}

\date{}

\section{Introduction}

The detection of cosmic neutrinos is a new and unique method to study the Universe. The weakly interacting nature of neutrinos makes them a complementary cosmic probe to other messengers such as the electromagnetic radiation, $\gamma$-rays, gravitational waves and charged cosmic rays. Neutrinos can travel cosmological distances, crossing regions with high matter or radiation field densities, without being absorbed. They allow the observation of the distant Universe and the interior of the astrophysical sources.

A milestone has been set with the first evidence of a cosmic signal of high-energy neutrinos \cite{ICE1} by the IceCube detector \cite{ICEDetector,ICE2}. 
The ANTARES telescope  \cite{Gen},  although much smaller than the IceCube detector, is the largest undersea neutrino telescope currently in operation.
One of its main goals is the search for astrophysical point-like neutrino sources. 
To this aim, the pointing accuracy of the detector is important and an evaluation of this performance is required.

The interaction of cosmic rays in the atmosphere produces downward-going muons  that can be recorded by underground,  underice or underwater experiments. Atmospheric muons represent a  large source of background for cosmic neutrino detection, but at the same time they can be used to calibrate the detector.
Due to absorption effects of cosmic rays by the Moon, a deficit in the atmospheric muon event density (expressed as number of events per square degrees)
in the direction of the Moon, the so-called {\em Moon shadow}, is expected.
With this approach, the Moon shadow has been already measured and reported by MACRO \cite{MAC}, SOUDAN \cite{SOU}, L3+Cosmics \cite{L3C} and  by IceCube \cite{ICE} Collaborations. It is worthy to mention here that other experiments, like  CYGNUS \cite{CYG}, TIBET \cite{TIB}, CASA \cite{CAS}, ARGO-YBJ \cite{ARG}, and recently also HAWC \cite{HAWK} measured the  Moon shadow by exploiting surface arrays detectors.

This work presents the first measurement of ANTARES angular resolution with atmospheric downward-going muons and the detector pointing performance making use of a celestial source for calibrations. A complementary estimation of the telescope pointing accuracy has been performed by  means of a {\it surface array} of particle detectors 
 arranged onboard a ship deck. The ship  was temporarily routing above the  ANTARES detector, allowing to  correlate the signals from  the detection of  atmospheric showers  with  the signals induced by downward-going muons in the underwater telescope.

This paper is organized as follows: in Section~\ref{sec:antares} the ANTARES detector is introduced together with the motivations of the present analysis; 
in Section~\ref{sec:moon_shadow_analysis} the  Moon shadow analysis is described; the surface array analysis is  presented in Section~\ref{sec:surface_array} and the conclusions are reported in Section~\ref{sec:conclusions}.

\section{The ANTARES neutrino telescope} \label{sec:antares}

The ANTARES detector is deployed 40 km offshore from Toulon, France ($42^\circ48'$N, $6^\circ10'$E) anchored at a depth of about 2475 m.  
The telescope measures the  Cherenkov light stimulated in the medium by  relativistic particles  by means of a three dimensional grid of optical modules (OMs), pressure resistant glass spheres each containing  one 10$^{\prime\prime}$ photomultiplier tube (PMT).
The OMs are arranged in triplets, forming a storey, along twelve vertical lines, for a total of 885 OMs \cite{Gen}.
The lines are anchored on the sea bottom and kept taut by a buoy at the top. Each PMT is nominally oriented $45^\circ$ downward with respect to the vertical. 
This orientation enhances the efficiency for the reconstruction of upward-going tracks, but still allows the detection of downward-going muons with smaller efficiency.
A titanium cylinder in each storey houses the electronics for readout and control, together with compasses and tiltmeters.
The total length of each line is 450 m, without any instrument along the lower 100 m. The distance between storeys is 14.5 m and the  distance between two lines ranges between 60 m and 75 m. The lines are connected to a central junction box which, in turn, is connected to shore via an electro-optical cable. Due to sea currents, 
a positioning system comprising hydrophones, compasses and tiltmeters is used to monitor the detector geometry \cite{tim}.
Finally the absolute orientation is provided by the triangulation 
of acoustic signals between lines and the deployment vessel at the sea surface 
using GPS \cite{tim} \cite{Cal}. The first detection line was deployed in 2006; the detector was completed in 2008.

The recorded information of each photon detected on a PMT
 is referred to as {\em hit}, 
and consists of the detection time, the amount of electric charge measured on the PMT anode and the PMT identification.
The ensemble of hits  contained in a certain time-window, identified   after some trigger condition, is called  {\em event}. 
Muon candidates are identified by requiring spacetime causality  between  the hits of one event \cite{BBfit}\cite{AAfit}. 
The quality of the reconstruction of  muon trajectories depends on  the goodness of such spacetime correlation.

\section{The Moon shadow analysis} \label{sec:moon_shadow_analysis}

Atmospheric muons are a valuable resource for validating the detector performance  and characterising some of the possible systematics associated to the experimental setup.
Muons   produced in the interactions of primary cosmic rays in the upper  layers of the atmosphere  can traverse several kilometres of water equivalent; for this reason only downward-going atmospheric muons can be measured  \cite{AGUILAR2010179,AGUILAR201086,ADRIANMARTINEZ201643}.
For those primary cosmic rays absorbed by the Moon, a  deficit  in the flux of the  secondary muons can be measured, being  directly correlated to the position of the Moon in the sky. 

\begin{figure}[htbp]
\begin{center}
\makebox[\textwidth][c]
{
\includegraphics[width =16cm]{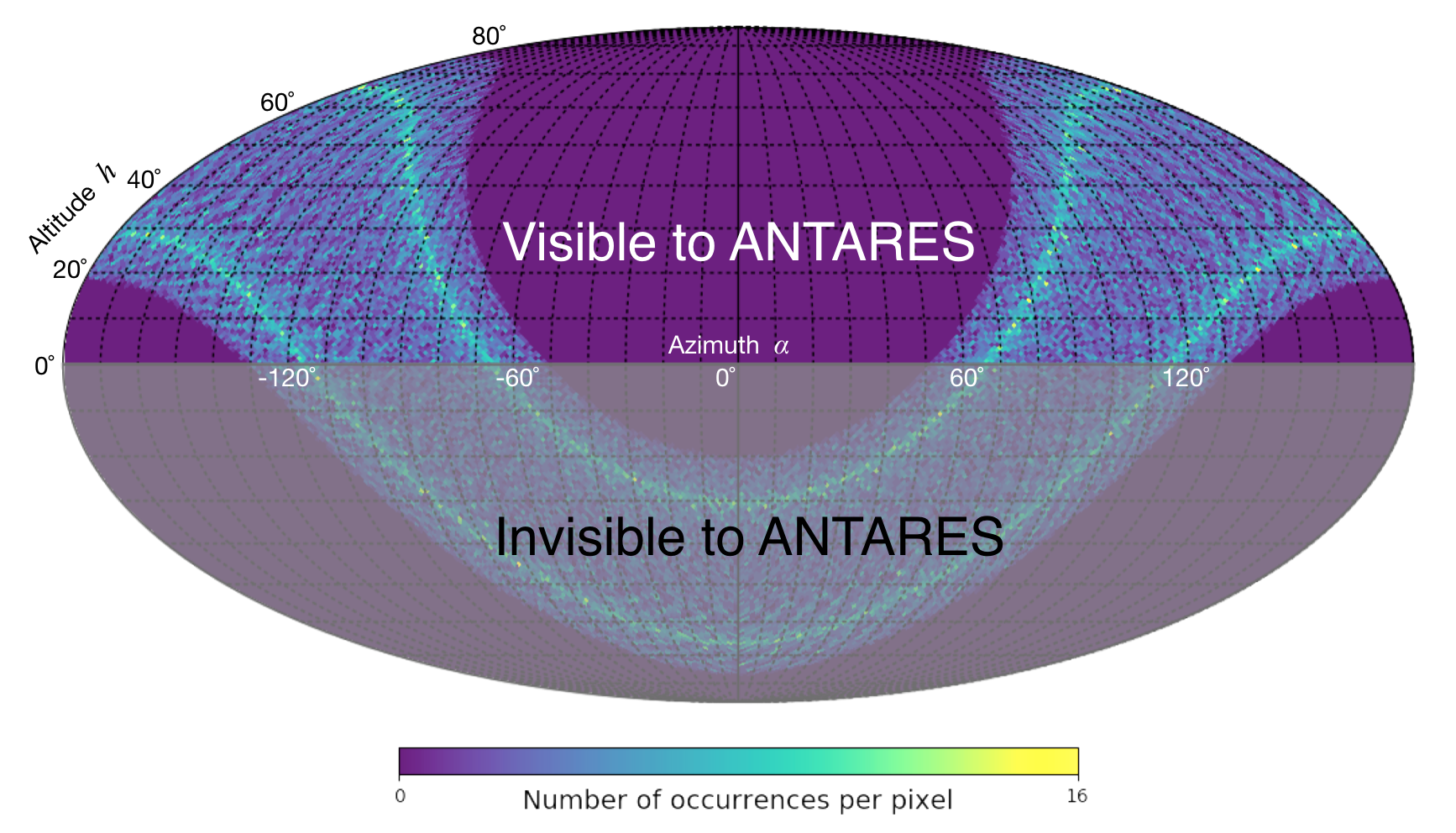}
}
\caption{The visible 
and invisible sectors 
of the position of the Moon for ANTARES with respect to the detector horizontal coordinate system. 
The occurrences of the Moon position are computed at each hour in the period 2007-2016 with the library {\em SkyField} \cite{SkyField}. The map is arranged according to the Mollweide equal-area view obtained with the use of the {\em HEALPIX} package \cite{Healpix}, setting the parameter $NSIDE = 64$ (\ie 49152 pixels).}
\label{fig:moon-position}
\end{center}
\end{figure}

The energy threshold for muons detectable at the depth of the ANTARES telescope is about 500 GeV when they are at the sea surface level, most of them with energy above 1 TeV.   Primaries which are progenitors of such highly energetic muons are practically not affected by the Earth geomagnetic field. 
This assumption of large rigidity   holds also for the secondary muons detected by the ANTARES detector,  thus they can be exploited in the study of the  Moon shadow without introducing any bias.
The smearing of muon direction with respect to the primary cosmic rays due to  pion transverse momentum  and pion decay  is limited by the large Lorentz factor \cite{Carla}.  
The analysis presented in this paper covers the  data-taking period spanning from 2007 to 2016, corresponding to a total live-time of 3128 days. Figure~\ref{fig:moon-position} shows the position of the Moon in the horizontal coordinate system of the detector for such a period.  
 The Moon altitude ranges above the horizon up to about $75^\circ$.

The analysis is performed in three steps, described in sub-sections \ref{sec:quality_cuts}, \ref{sec:1Danalysis} and \ref{sec:absolute_pointing}.
First, quality cuts are defined  to reduce
the number of candidate atmospheric muon events to a sample which provides the best sensitivity for this search.
The second part concerns the estimation of the telescope angular resolution for atmospheric muons by studying the mono-dimensional profile of the Moon shadow. In the third part, the pointing precision is determined evaluating a possible shift of the measured direction of the Moon with respect to the nominal values provided by astronomical libraries \cite{AstroLibMeeus}.
 
\subsection {Optimisation of quality cuts}\label{sec:quality_cuts}

The selection criteria applied to the reconstructed muon tracks are optimized using a dedicated Monte Carlo (MC) production. 
The MC generation of the atmospheric muon sample is performed with the MUPAGE code \cite{Mupage}. MUPAGE implements parametric formulas for the flux, the radial distribution, the multiplicity and the energy spectrum of muons at a given depth, allowing for a
fast production of both single and bundle muon events. Muons are generated on the surface of a cylinder-shaped volume of water, 650 m high, with a radius of 290~m, containing the detector.
This volume is larger than the instrumented volume and corresponds to the region in which muons can  produce detectable signals.
The generation of the MC sample is subdivided in different batches corresponding to the  periods of data-taking, referred to as $runs$.  The simulation reproduces the effective data taking conditions of the ANTARES detector, which can vary on a run-by-run basis \cite{FUSCO}.
The simulation includes the generation of Cherenkov light stimulated by the muon  and its propagation up to the PMTs on the basis of the  measured characteristics of light  propagation \cite{sito}.
Optical background, caused by bioluminescence and radioactive isotopes (mainly $^{{40}}$K) present in sea water, is also added according to the measured rate.
This technique allows to correlate the actual  time of each run to the  position of the  Moon in the sky. 
In particular, it is possible to assign an absolute time-stamp,  generated randomly within the period of each considered run, to each MC  event reconstructed as a downward-going muon.
A detailed production compliant with the actual live time  is used to generate, reconstruct and select the MC sample of events within the restricted area of 10$^{\circ}$ around the nominal position of the Moon at the time of each event.  
In order to evaluate the contamination of mis-reconstructed events in the proximity of the Moon, a smaller MC sample,  with 1/3 of the actual live time, is generated over the whole visible sky.

The detector response is then simulated taking into account the main features of the PMTs and of the electronics \cite{Ju,Margy}. 
Finally, the PMT signals are processed to reconstruct the atmospheric muon tracks with the standard ANTARES algorithm for track-like events. This is a robust track-fitting procedure based on a  likelihood maximisation  \cite{AAfit}. Figures of merit are determined by means of  two quality parameters:   $\Lambda$,  which varies linearly with the logarithm of the reconstructed track likelihood,
  and  $\beta$, the angular error associated to the reconstructed direction.

Two different MC simulation sets are prepared: the sample $S_{1}$ considering the shadowing effect of the Moon and the sample $S_{0}$ without this effect. In the sample $S_{1}$, the Moon shadow is obtained by removing the muons generated within the Moon disk, assuming a radius of 0.26$^\circ$. 
The information from all the considered simulated runs is combined to obtain  statistical evidence of the Moon shadow. 
For each of the two MC samples, $S_{1}$ and $S_{0}$, a one dimensional  histogram is built with the distribution of 
events as a function of the angular distance $\delta$ with respect to the Moon, up to 10$^{\circ}$. Such a histogram is subdivided into 25 bins, each one sized $\Delta \delta=0.4^{\circ}$ and   corresponding to an annulus of increasing radius centered on the Moon. The content of each bin is  normalised to the corresponding  annulus area, resulting  in an  event density.

The  cuts on the quality parameters $\Lambda$ and $\beta$ are chosen to achieve the best sensitivity for the Moon shadow detection. The approach of the hypothesis test is used: 
 the  null hypothesis $H_{0}$ relates to the case of atmospheric muons  without  the Moon shadow, while the alternative hypothesis $H_{1}$ corresponds to the presence of the Moon.  The used test statistic is  defined as  $\lambda=-2\log{\frac{L_{H_{1}}}{L_{H_{0}}}}$, with $L_{H_{0}}$ and $L_{H_{1}}$  the likelihoods obtained under the $H_{0}$ and $H_{1}$ hypotheses. Assuming that the event population in each bin follows a Poisson probability distribution, using the $\chi^2$ definition in \cite{PDG},  the chosen test statistic  can be conveniently written as:
\begin{eqnarray}
	\lambda&=&\chi^{2}_{H_{1}}-\chi^{2}_{H_{0}}\label{eq:lambda-delta-chi2} \\ 
	&\mbox{with}&\nonumber\\
	\chi^{2}_{H}&=&2\;\sum_{i=1}^{N_{bin}}\left[N_{i,H}-n_{i}+n_{i}\ln{\frac{n_{i}}{N_{i,H}}}\right], \label{eq:chi2}	
\end{eqnarray}
where $n_{i}$ stands for the  measurement in the $i$-th bin to be compared with the expectations $N_{i,H}$ under the  $H_{0}$ and $H_{1}$ hypotheses. 
The following reduced expression for $\lambda$ is  used:
\begin{eqnarray}
	\lambda&=& 2\;\sum_{i=1}^{N_{bin}}\left[\mu_{i}-\nu_{i}+n_{i}\ln{\frac{\nu_{i}}{\mu_{i}}}\right],
	\label{eq:lambda-test-statistics}
\end{eqnarray}
where for simplicity   the expected counts   $N_{i,H_{0}}$ and $N_{i,H_{1}}$ are renamed as  $\nu_{i}$ and $\mu_{i}$, respectively.
The two possible  distributions of $\lambda$, $f(\lambda|H_{0})$ and $f(\lambda|H_{1})$, valid separately under the hypotheses $H_{0}$  and $H_{1}$, respectively, are obtained by means of  pseudo-experiments (PEs). The   number of events in the $i$-th bin $n_{i}$  is determined  by extracting $10^{6}$ random  values generated according to a Poisson distribution with expectation values equal to $\nu_{i}$ and $\mu_{i}$.

\begin{figure}[htbp]
\begin{center}
\includegraphics[width =9.5cm]{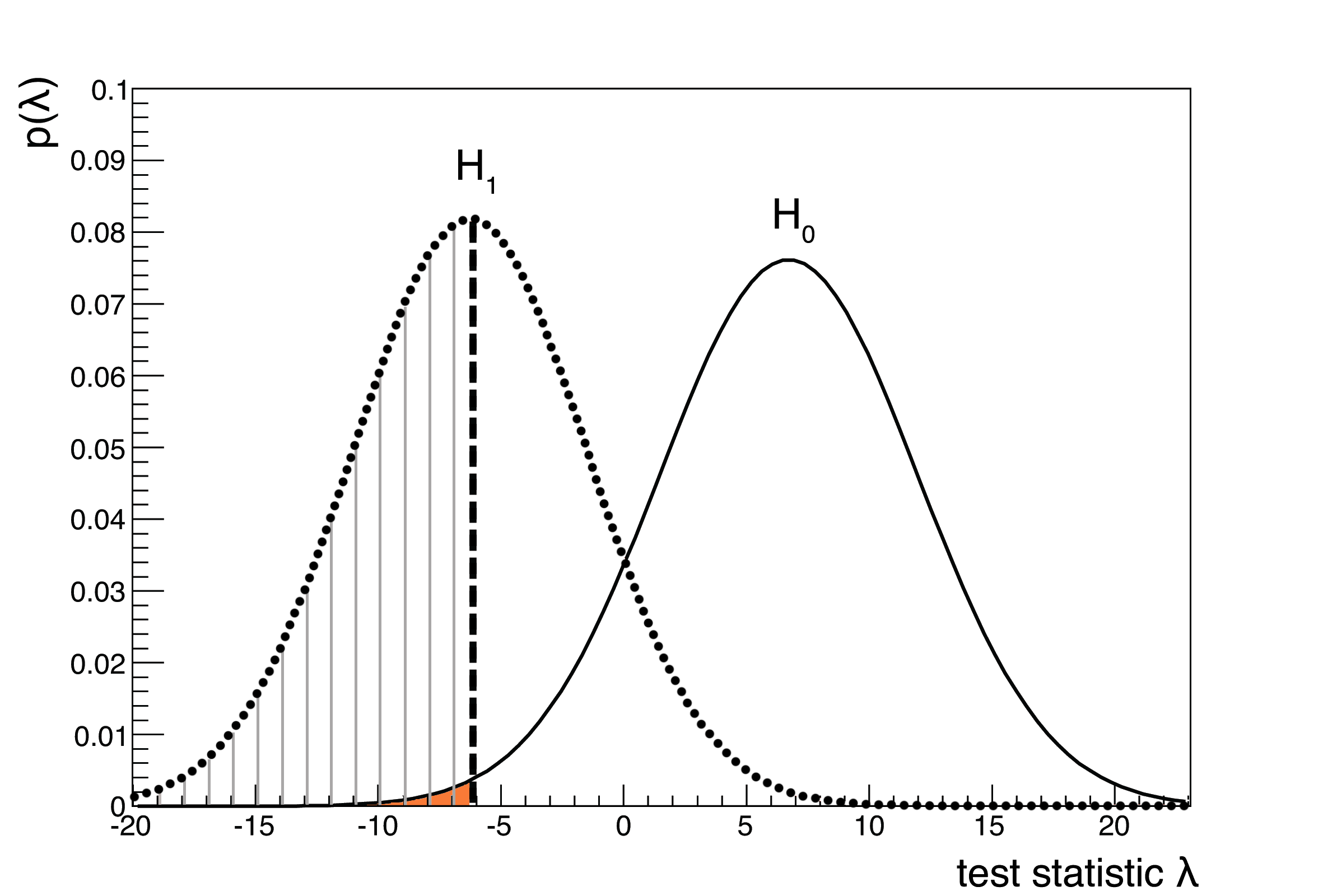}
\caption{The test statistics $\lambda$ distribution for the ``Moon shadow'' hypothesis $H_{1}$
(dotted curve) and  the ``no Moon shadow'' hypothesis $H_{0}$ (smooth curve). The dashed area corresponds to the 50\% of the pseudo-experiments where the  Moon shadow hypothesis is correctly identified. The shaded area quantifies the expected median significance (here $3.4\,\sigma$) to observe the Moon shadow.}
\label{fig:distr}
\end{center}
\end{figure}

Several hypothesis tests are performed assuming different selection criteria for $\Lambda$ and $\beta$. For each set of values, the distributions $f(\lambda|H_{0})$ and $f(\lambda|H_{1})$ are compared. 
The median of $f(\lambda | H_{1})$ is taken as the critical value for $\lambda$, \ie as the threshold to separate the two hypothesis.
The set of best cut values of $\Lambda$ and $\beta$ corresponds to that for which the two $f(\lambda|H)$  distributions have the minimal overlap.
Figure~\ref{fig:distr}  shows the distribution $f(\lambda|H_{0})$ (black curve) and $f(\lambda|H_{1})$ (red curve)  for the optimised quality cuts $\Lambda_{cut}= -5.9 $,  $\beta_{cut}= 0.8^{\circ}$, and the critical value is  $\lambda=-6.15$.
The dashed area below $f(\lambda|H_{1})$ represents the fraction of PEs where the Moon shadow hypothesis is correctly identified; the filled-coloured area below $f(\lambda|H_{0})$ corresponds to a \pvalue equal to $3.6\times10^{-4}$, or equivalently $3.4\,\sigma$. This is the  expected median significance of the Moon shadow effect with the MC data set.

\subsection{Deficit significance and angular resolution} \label{sec:1Danalysis}
\begin{figure}[htbp]
\begin{center}
\includegraphics[width=320pt, keepaspectratio=true]{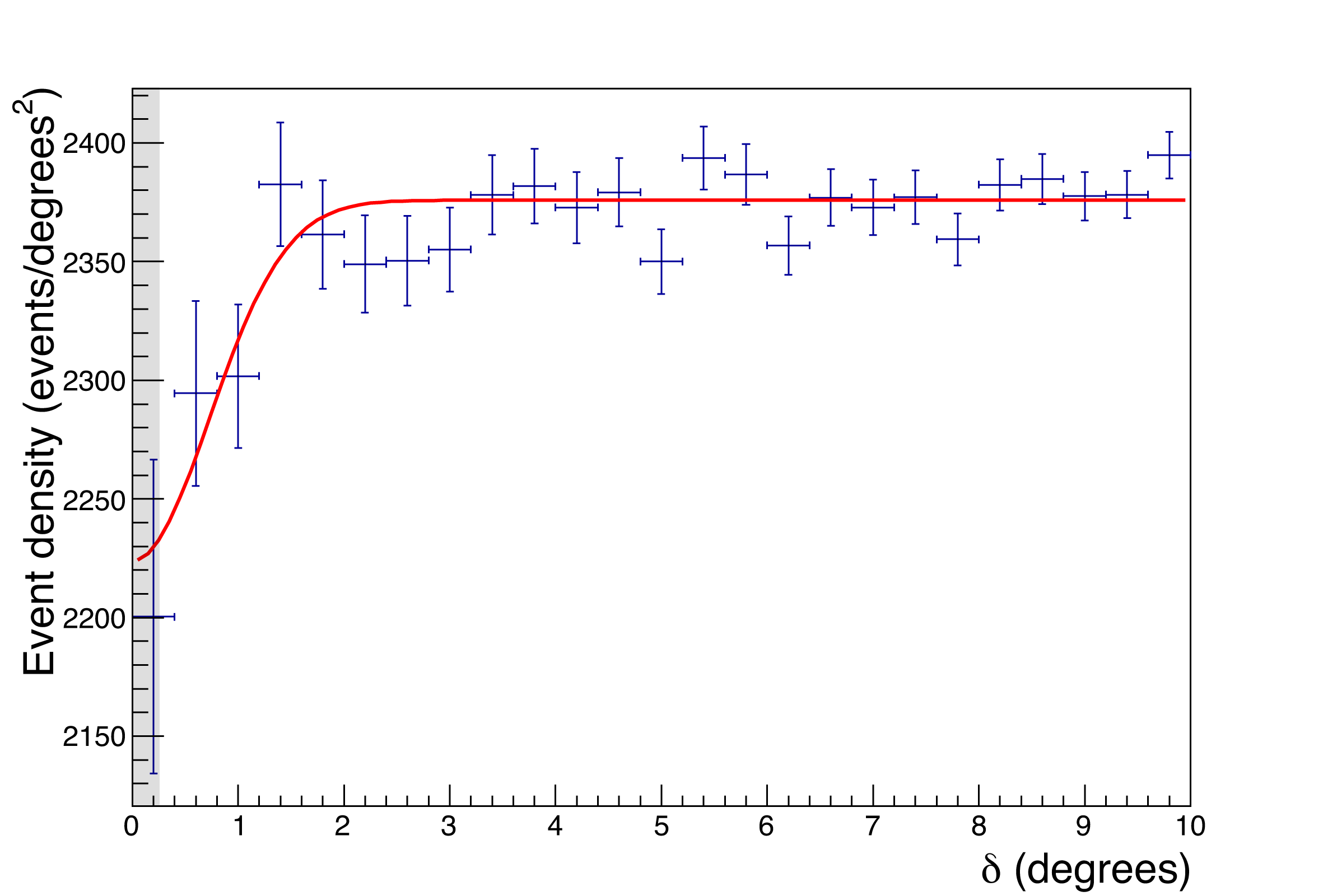}
\caption{
Measured muon event density as a function of the angular distance $\delta$ from the Moon. Data histogram is shown with statistical errors; the smooth line is the best fit according to equation (\ref{eq:fit}); the shaded area corresponds to the apparent radius of the Moon (0.26$^\circ$).}
\label{fig:datagaussfit}
\end{center}
\end{figure}

The  optimized quality cuts  reported above are applied to the data sample collected in the period 2007-2016.
Figure~\ref{fig:datagaussfit} shows the resulting distribution for the  muon density as a function of  $\delta$ in the range of $0^{\circ}\leq\delta\leq 10^{\circ}$ with bin size of $\Delta\delta=0.4^{\circ}$. 
A clear deficit of events is evident in the region around the Moon position ($\delta<1.2^\circ$).

For the estimation of  the  angular resolution,  the Moon shadowing effect is assumed to follow a Gaussian distribution with standard deviation $\sigma_{res}$, which corresponds to  the detector angular resolution itself. This is motivated by the fact that the apparent size of the Moon in the sky is sufficiently small compared to the expected value of the detector angular resolution, affecting the estimation by less  than a few percents.  
 A similar approach has already been followed by \cite{MAC} \cite{ICE} \cite{MINOS}.
The number of expected events is evaluated by fitting the distribution in Figure~\ref{fig:datagaussfit} with the following function \cite{SOU}:
\begin{center}
\begin{equation}
  \frac{dn}{d\delta^2}=k({1-\frac{R^2_{Moon}}{2\sigma_{res}^2}e^{-\frac{\delta^2}{2\sigma_{res}^2}}}) .
\label{eq:fit}
\end{equation}
\end{center}
The two free parameters are $k$, the average muon event density in the $H_{0}$ scenario, and  $\sigma_{res}$. The Moon radius $R_{Moon}$ is fixed to 0.26$^\circ$.

The angular resolution for downward-going atmospheric muons resulting from the fit is $\sigma_{res} = 0.73^\circ$$\pm$0.14$^\circ$, with the fitted value of $k = 2376\pm$3 events per square  degrees.  The  goodness of the fit is found to be  $\chi^{2}/\mbox{dof} = 23.5/23$.

The significance of the shadowing is evaluated using a $\chi^2$~test comparing the measured event density with the flat distribution $\frac{dn}{d\delta^2}=k$. 
Such $\chi^2$ test leads to a \pvalue equal to $4.3\times10^{-4}$ corresponding to a  significance of the Moon shadow effect of  $3.3\,\sigma$. 
This value is compatible with the expected significance of the Monte Carlo  previously described.

\subsection{Absolute pointing}\label{sec:absolute_pointing}
The procedure for evaluating the  pointing accuracy  of the Moon shadow is partially inspired by \cite{MAC}; it is based on determining the statistical significance of the selected data set under the assumption of the Moon  in a given direction. All possible placements  are considered within a field of view (FoV) centered on the nominal position of the Moon. This work differs from \cite{MAC} in  the way the significance of the results is  evaluated.
 
The event distribution of the detected muons, compliant to the determined quality cuts, is represented as  function of $x=(\alpha_{\mu}-\alpha_{Moon})\times\cos(h_{\mu})$ and $y=h_{\mu}-h_{Moon}$; here $(\alpha_{\mu},h_{\mu})$ and $(\alpha_{Moon},h_{Moon})$ are the horizontal coordinates of the track and the Moon, respectively, at the time of the event.
The FoV is limited in both $x$ and $y$ within the range $\left[-10^{\circ},10^{\circ}\right]$, and it is subdivided in a grid of 0.2$^{\circ}\times0.2^{\circ}$ squared bins.
The used test statistic is again $\lambda$ as reported in equation (\ref{eq:lambda-test-statistics}), but now the sum is evaluated on all $100\times100$ square bins.

The expectations under $H_{0}$ are obtained parameterising  the event distribution of the measured  atmospheric muons which fall in the FoV relative to  the position of the Moon  four hours before the timestamp of each event. 
The  parameterisation is done with a second degree polynomial of the form:
\begin{eqnarray}
	p_{2}(x,y,\vec{k}|H_{0})= k_{0}+k_{1}x+k_{2}x^{2}+k_{3}y+k_{4}y^{2}, \label{eq:poly-bkg}
\end{eqnarray}
with the fitted parameter array $\vec{k} \equiv\{ 93.6 \pm 1.8, 0.19 \pm 0.16,(-8.2 \pm 3.1)\times 10^{-3},$ $ 3.98 \pm 0.17, (5.60 \pm 0.32)\times 10^{-2}\}$.
The goodness of the fit for this set of values $\vec{k}$ is $\chi^{2}/\mbox{dof} = 9993/9995$, corresponding to a \pvalue $\approx$ 0.5; it validates the  modelling of the event distribution in the absence of the Moon provided by equation (\ref{eq:poly-bkg}).  

\begin{figure}[h!]
\centering
\begin{subfigure}[b]{0.5\columnwidth}
\centering
\includegraphics[width=\linewidth,keepaspectratio]{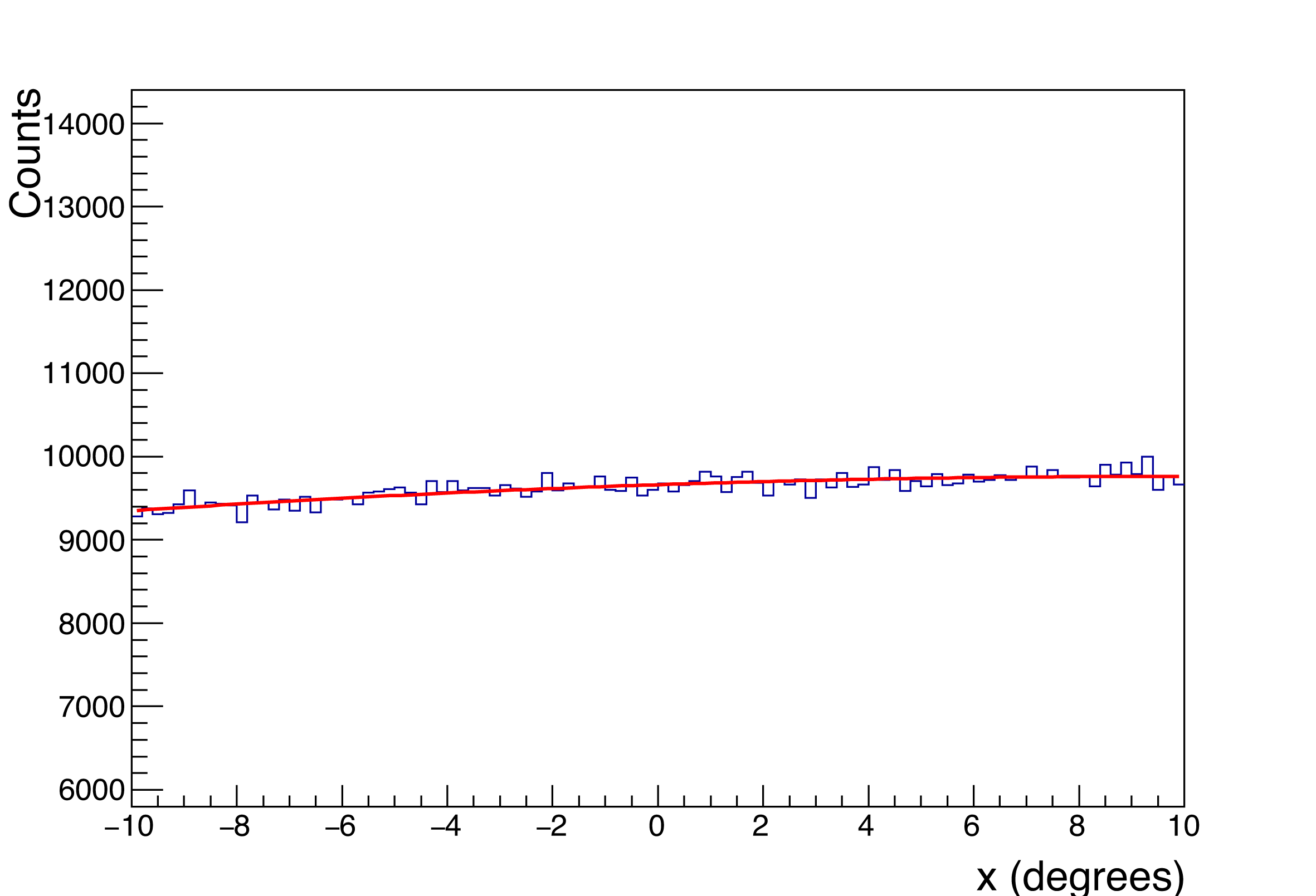}
\caption{}
\label{fig:marginal_x}
\end{subfigure}\hfill
\begin{subfigure}[b]{0.5\columnwidth}
\centering
\includegraphics[width=\linewidth,keepaspectratio]{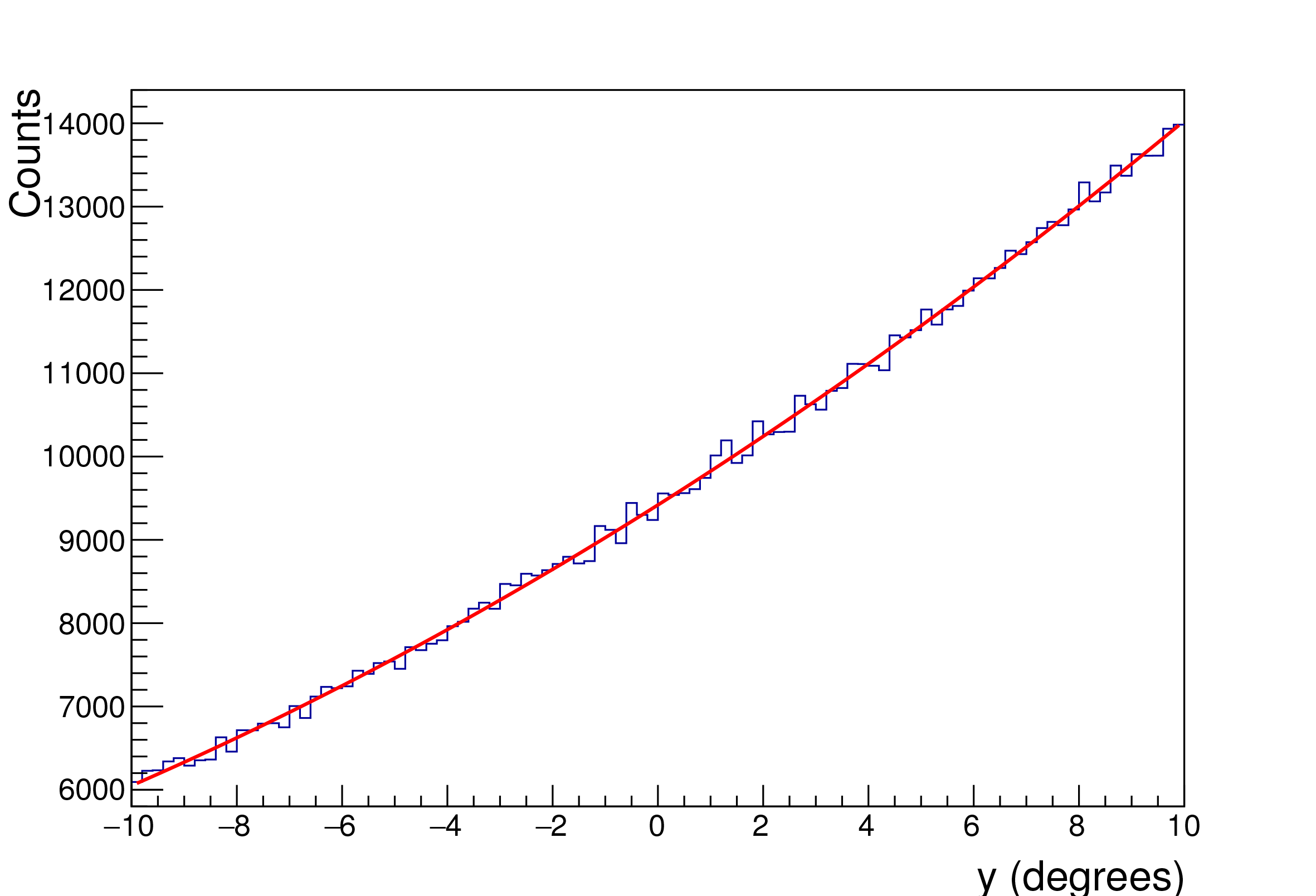}
\caption{}
\label{fig:marginal_y}
\end{subfigure}
\caption{Projection of the measured 2-D event distributions, in absence of the Moon shadow ($H_{0}$), for the field of view coordinates $x  = (\alpha_{\mu}-\alpha_{Moon}) \times \cos{h_{\mu}}$ and $y = h_{\mu}-h_{Moon}$. 
}
\end{figure}

Figures \ref{fig:marginal_x} and \ref{fig:marginal_y} represent the projection of the event distribution in the FoV onto the  $x$ and $y$ axes  in the absence of the Moon shadow, also called {\it marginal distributions}. The marginal distribution for $x$  is almost flat, compliant with the expected lack of  any significant structure in the atmospheric muon flux along the azimuth. On the contrary, the marginal distribution for $y$ shows an almost linear ramping which reflects the enhancement of the muon flux with the altitude.

The expectations under $H_{1}$ are then obtained by  subtracting  from 
  $p_{2}(x,y,\vec{k}|H_{0})$ a bi-dimensional Gaussian point spread function: 
\begin{eqnarray}
	G(x,y,\vec{\theta})&=& \frac{A}{2\pi\sigma^{2}_{res}}\,e^{-\frac{(x-x_{s})^{2}+(y-y_{s})^{2}}{2\sigma^{2}_{res}}}\label{eq:gaussian-psf}.
\end{eqnarray} 
In equation (\ref{eq:gaussian-psf}) the same spread is assumed in both dimensions, so that $\sigma_{x}=\sigma_{y}\equiv\sigma_{res}$. The $\sigma_{res}$  is fixed  to value of the angular resolution found in the previous sub-section.  The array of free parameters   $\vec{\theta}$  is composed  of  the amplitude of the Moon deficit $A$ and the assumed position of the Moon $(x_{s},y_{s})$ in the FoV.

For each bin in the FoV, the value of the test statistic $\lambda$ is minimised finding the best estimation of $A$. The smallest value  $\lambda_{min}$ is found  equal to \lshift, for the fitted deficit amplitude  $A_{min}=\;$\AMshift\AMshifterr, in the bin with  center in $x=0.5^{\circ}$ and $y=0.1^{\circ}$. Such coordinates   are taken as the best estimation of the position of the Moon. The test statistic $\lambda_{O}$ in the nominal position $O\equiv( 0^{\circ},0^{\circ} )$  is found equal to \lnom for the corresponding amplitude $A_{O}=$\AMnom\AMnomerr.
At each bin, $-\lambda$ follows the  distribution of a  central $\chi^{2}$ with one degree of freedom,
 assuming $H_{0}$ as true. This allows to estimate the discrepancy of  the measured data  from the assumption of the absence of the Moon. Considering $\lambda_{O}$,  a \pvalue of \pvHzero is obtained, which corresponds to a  significance of \soneHzero, in agreement with what is reported in the above section \ref{sec:1Danalysis}.

\begin{figure}[htbp]
\begin{center}
   \includegraphics[width=.7\linewidth]{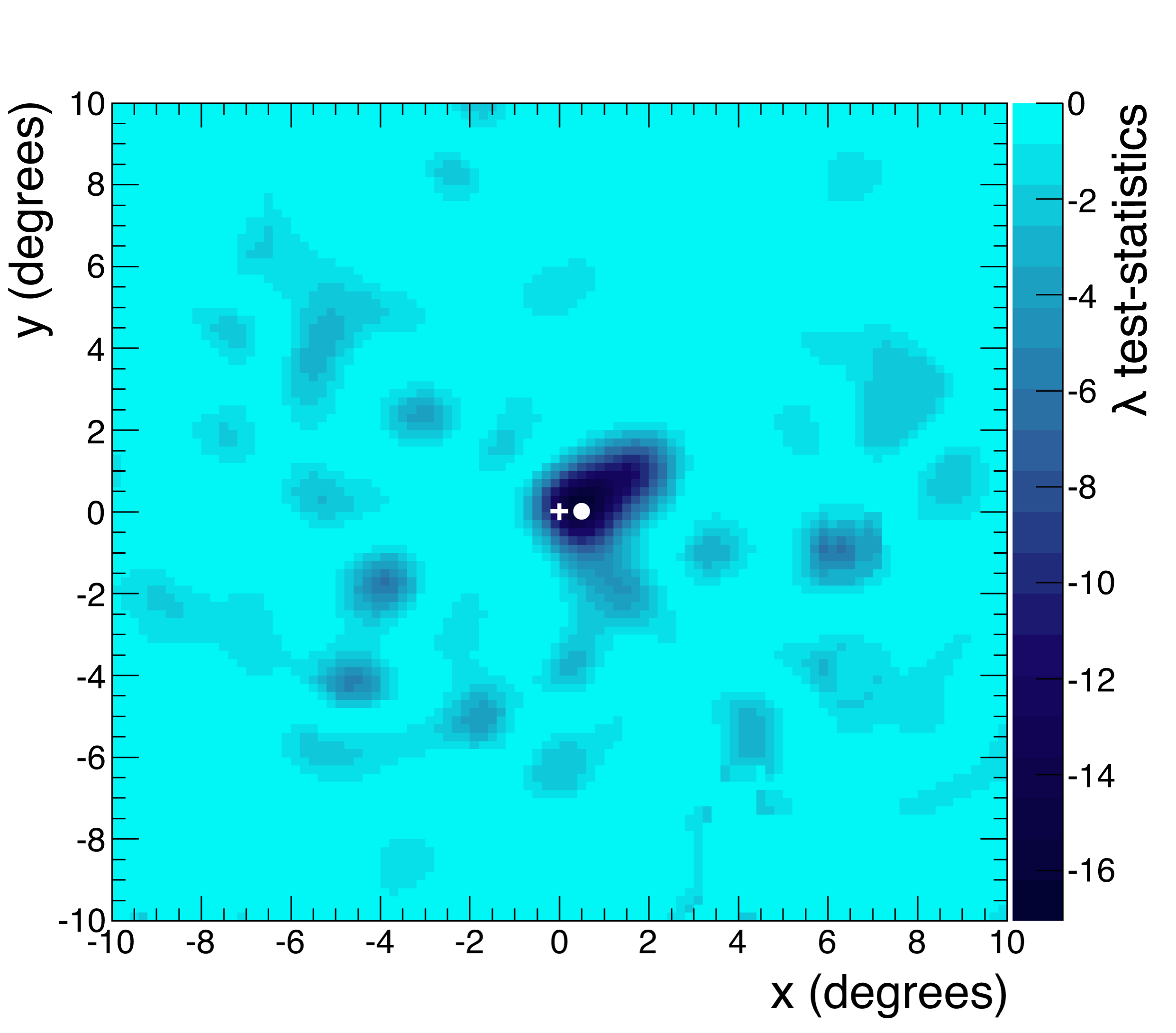}
   \caption{ Measured distribution of the test statistic $\lambda$ from equation (\ref{eq:lambda-test-statistics}) in the field of view around the Moon nominal position  $O\equiv(0^{\circ},0^{\circ})$, indicated by a white cross. The white dot refers to the coordinates $(0.5^{\circ},0.1^{\circ})$ where the test statistics reaches the minimum ($\lambda_{min}=$\lshift). }
   \label{fig:lambda_map} 
\end{center}
\end{figure}

Figure~\ref{fig:lambda_map} shows the $\lambda$ distribution in the FoV. It  can be interpreted as a bi-dimensional profile-likelihood, with $A$ treated as the nuisance parameter.
The interval  corresponding to a desired confidence level ($CL$)  is obtained for $\lambda\leq\lambda_{cut}=\lambda_{min}+Q$, where $Q$ is the quantile accounting for two degrees of freedom  and confidence level $CL$ \cite{COWAN}.   

An additional strategy is used to cross-check the confidence intervals found with the method reported above. 
This is done by exploiting the PE technique. In each bin of the FoV, a reference number of events $\{n_{i}\}_{ref}$ is computed using the superposition of  equations (\ref{eq:poly-bkg}) and (\ref{eq:gaussian-psf}). For this purpose  the Moon is assumed to be in $O$, $\sigma_{res}=0.73^{\circ}$ and $A\;=\; A_{O}$. For each PE, a corresponding data set $\{n_{i}\}_{PE}$ is extracted  as  Possionian fluctuations of  the reference set $\{n_{i}\}_{ref}$.  Using $10^{5}$ PEs, the distribution of the best value of $\lambda_j$ is determined at the $j$-th bin of the FoV.
For each $\lambda_{j}$ distribution, the range $\left(-\infty,\lambda^{CL}_{j}\right]$ is considered, where $\lambda^{CL}_{j}$ is the value of $\lambda_{j}$ such that its cumulative distribution is $F(\lambda^{CL}_{j})=CL$; the $j$-th bin is included  into the  confidence interval if 
 $\lambda^{m}_{j}\leq\lambda^{CL}_{j}$. 

\begin{figure}[htbp]
\begin{center}
	\includegraphics[width=.7\linewidth]{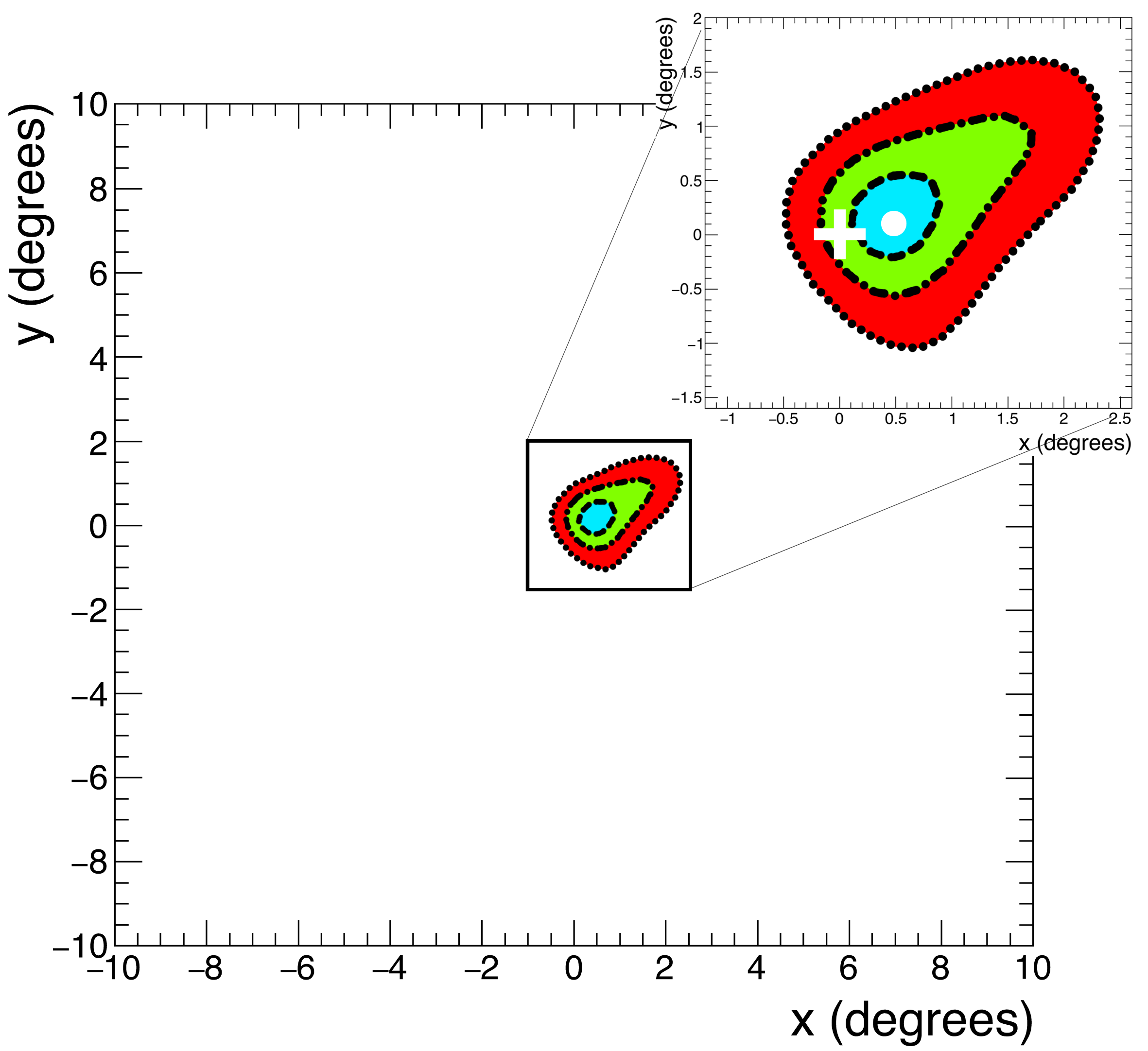}
   \caption{Contour plots corresponding to different confidence levels  (cyan/dashed: 68.27\% ; green/dot-dashed: 95.45\% ; red/dotted: 99.73\% ), computed with the two methods described in the text. In the zoom, the  dot represents the position in the FoV where $\lambda_{min}=$\lshift. The  cross indicates the nominal position of the Moon.}
   \label{fig:contour_combo}
\end{center}
\end{figure}

Figure~\ref{fig:contour_combo} shows the estimation of the confidence regions for~$CL\equiv\{$68.27\%, 95.45\%, 99.73\%$\}$ using both the methods explained above. 
The contours found with the first and the second methods are indicated by colours and lines, respectively.
The contour plots of the two approaches are in excellent agreement.

\begin{figure}[htbp]
\begin{center}
\includegraphics[width=360pt, keepaspectratio=true]{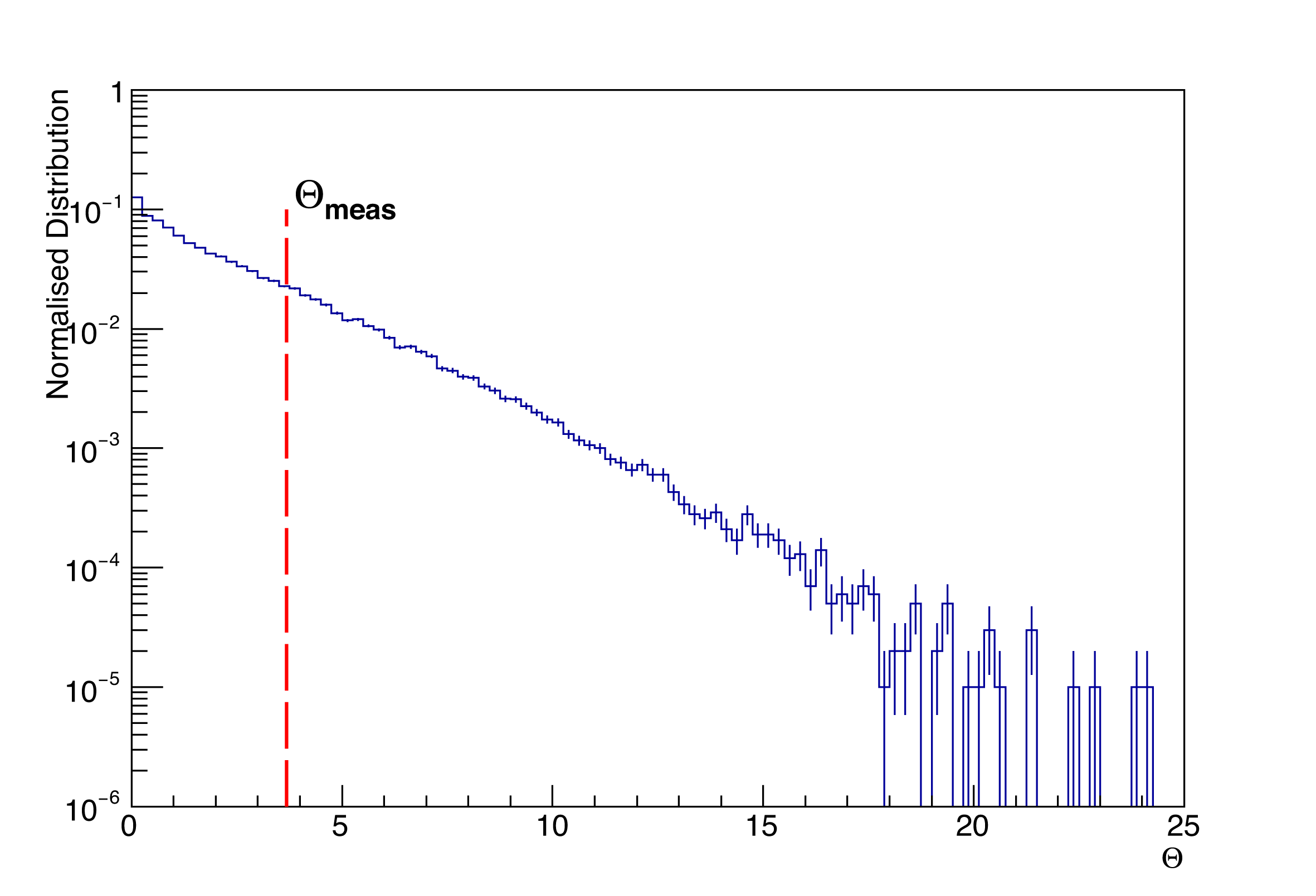}
\caption{Distribution of the $\Theta=\lambda_{O}-\lambda_{min}$ test statistics, obtained with $10^{5}$ pseudo-experiments  assuming the Moon in the nominal position $O$. The 23\% of the pseudo-experiments has a test statistic $\Theta$  larger than the measured value $\Theta_{meas}=3.68$, indicated for reference by the red-dashed line}.
\label{fig:delta_lambda_distrib}
\end{center}
\end{figure}

The statistical significance of the apparent shift with respect to the Moon nominal position is determined using PEs. The method relies on the  probability density function of the test statistics
 $\Theta=\lambda_{O}-\lambda_{min}$, with  $\lambda_{O}$ and $\lambda_{min}$  defined as before. 
 The $\Theta$ test statistic is interpreted as a profile likelihood whose distribution asymptotically tends to 
 that one of a $\chi^{2}$ with two degrees of freedom. In Figure \ref{fig:delta_lambda_distrib} the normalised distribution of 
 the $\Theta$ test statistic is shown, where the measured value of the test statistic $\Theta_{meas}=\;$\dlmeas is indicated for reference by the  red-dashed line. 
Integrating the $\Theta$ distribution for values larger than $\Theta_{meas}$, a \pvalue = \pvdelta is obtained,  corresponding to a \stwodelta significance. This indicates that the shift is compatible with a statistical fluctuation.

\section{Analysis of data collected with a surface array} \label{sec:surface_array}

The pointing performance of the ANTARES telescope is cross-checked  in a completely independent way, exploiting the measurements made with a surface array detector.
  The device was temporarily  onboard of a ship circulating around the position of the telescope, synchronised to a GPS reference. 
  The surface array was composed of a set of 15 liquid scintillator detection  units, designed for the measurement of atmospheric showers, 
  placed over an area of about 50 m $\times$ 14 m on the ship deck. 
  \begin{figure}[ht]
\begin{center}
\includegraphics[width=423pt, height=175pt, keepaspectratio=true]{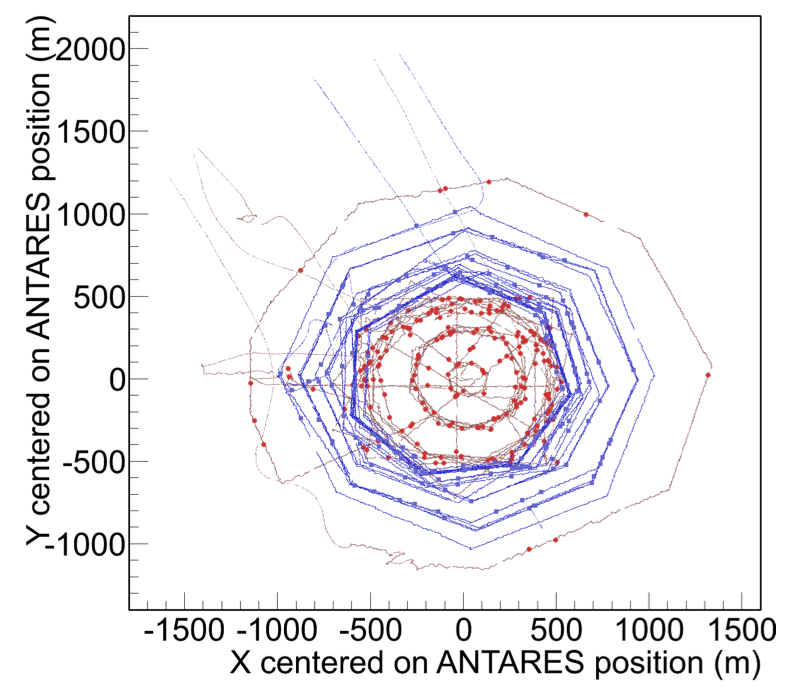}
\caption{Ship route in the 2011 (red) and 2012 (blue) campaigns around the center of the ANTARES detector.}
\label{fig:boat_arr}
\end{center}
\end{figure}
Each scintillator unit included a  polyethylene-aluminium box filled with linear alkylbenzene doped with  wavelength shifters. 
The scintillation light was detected using 2$^{\prime\prime}$ PMTs, one per unit.
Each scintillator unit had a single rate of around 100 Hz. The pointing accuracy of the ANTARES detector is inferred by combining the data from the surface array and the undersea telescope.

Two different sea campaigns were  performed: a first campaign of seven days in 2011 and a second campaign of six days in 2012.
Given the area covered by the ship routing above the  ANTARES telescope, the range of the  muon zenith $\theta$  is limited  to  $2^{\circ}\leq \theta \leq27^{\circ}$.
Figure~\ref{fig:boat_arr} shows the recorded positions on the sea surface of the ship 
during these two periods.

The shower array is used to trigger the possible time-correlations with the ANTARES events.
The typical trigger rate of the surface array is around 1 Hz requiring coincidences in at least 3 detection units in a 650~ns time window. 
The rate of reconstructed muons  is $\sim$ 0.25~Hz when applying cuts on the quality parameters $\Lambda\geq-6$ and $\beta\leq0.6^\circ$. 
The  coincidence time-window between the surface array and the underwater telescope is set to 10 $\mu$s. The rate of coincidences is about 40 per day, with  an expected rate of random coincidences of about 0.2 events per day.

\begin{figure}[htp]
\begin{center}
\includegraphics[width=323pt, height=175pt, keepaspectratio=true]{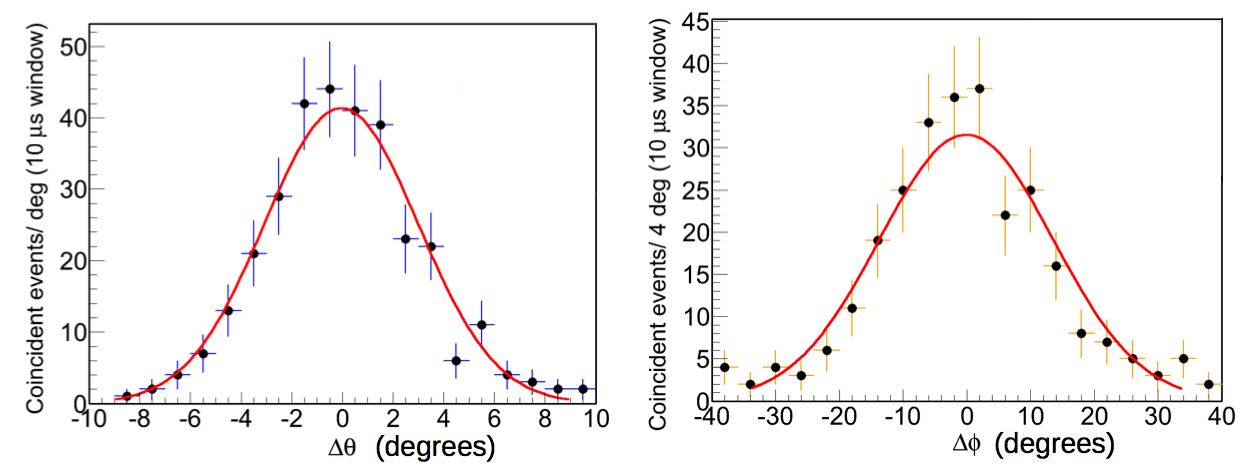}
\caption{Left: the difference between the zenith angles of the shower axis (determined as the direction of the ship with respect to the ANTARES location) and of the reconstructed muon underwater. Right: the same for the azimuth angles.  Fit results: $\Delta\theta_{mean}=-0.07^{\circ}\pm0.22^{\circ}$; $\Delta\phi_{mean}=-0.5^{\circ}\pm0.8^{\circ}$.}
\label{fig:surf_arr}
\end{center}
\end{figure}

The direction of the atmospheric shower is estimated by correlating the GPS position of the ship with the ANTARES location.  
 An uncertainty of 25 m, about one half of the ship deck hosting the shower  array, is assumed for the shower position detected by the array.  An error of 70 m is considered  for the possible displacement of the muon/muon bundle axis inside the detector volume. 
Considering only the ship routes with radius larger than 500 m, the estimated resolutions  are  $\sim$3$^\circ$ in zenith and $\sim$8$^\circ$ in azimuth.

The results of the two campaigns are shown in Figure~\ref{fig:surf_arr}: the represented  $\Delta\theta$ (left) and  $\Delta\phi$ (right)
are the  differences between the directions of the  shower axis and the reconstructed muon underwater.

According to the Gaussian fit of the two  distributions in Figure~\ref{fig:surf_arr}, the absolute pointing appears to be consistent with the nominal expectations, i.e. with a null systematic shift in both zenith and azimuth
within the errors (see caption of Figure ~\ref{fig:surf_arr}).
The large uncertainty in the azimuth estimation is due to the low zenith angle tested during the campaigns, as can be derived from Figure~\ref{fig:boat_arr}.
The results of the surface array analysis are in good agreement with the pointing performance found with the Moon shadow analysis.

\section{Conclusions} \label{sec:conclusions}

This paper describes the estimations of the pointing performance of the ANTARES telescope using the Moon shadow effect and a dedicated surface array.

The selected events from the data recorded in the  2007-2016 period with altitude angles $0^{\circ}\leq h\leq75^{\circ}$, allowed the identification of the Moon shadow with  \soneHzero~\ statistical significance. The corresponding detector angular resolution for downward-going atmospheric muons is  0.73$^\circ$$\pm$0.14$^\circ$.

The pointing accuracy of the detector is consistent with the expectations.

\section*{Acknowledgements}
The authors acknowledge the financial support of the funding agencies:
Centre National de la Recherche Scientifique (CNRS), Commissariat \`a
l'\'ener\-gie atomique et aux \'energies alternatives (CEA),
Commission Europ\'eenne (FEDER fund and Marie Curie Program),
Institut Universitaire de France (IUF), IdEx program and UnivEarthS
Labex program at Sorbonne Paris Cit\'e (ANR-10-LABX-0023 and
ANR-11-IDEX-0005-02), Labex OCEVU (ANR-11-LABX-0060) and the
A*MIDEX project (ANR-11-IDEX-0001-02),
R\'egion \^Ile-de-France (DIM-ACAV), R\'egion
Alsace (contrat CPER), R\'egion Provence-Alpes-C\^ote d'Azur,
D\'e\-par\-tement du Var and Ville de La
Seyne-sur-Mer, France;
Bundesministerium f\"ur Bildung und Forschung
(BMBF), Germany; 
Istituto Nazionale di Fisica Nucleare (INFN), Italy;
Nederlandse organisatie voor Wetenschappelijk Onderzoek (NWO), the Netherlands;
Council of the President of the Russian Federation for young
scientists and leading scientific schools supporting grants, Russia;
National Authority for Scientific Research (ANCS), Romania;
Mi\-nis\-te\-rio de Econom\'{\i}a y Competitividad (MINECO):
Plan Estatal de Investigaci\'{o}n (refs. FPA2015-65150-C3-1-P, -2-P and -3-P, (MINECO/FEDER)), Severo Ochoa Centre of Excellence and MultiDark Consolider (MINECO), and Prometeo and Grisol\'{i}a programs (Generalitat
Valenciana), Spain; 
Ministry of Higher Education, Scientific Research and Professional Training, Morocco.
We also acknowledge the technical support of Ifremer, AIM and Foselev Marine
for the sea operation and the CC-IN2P3 for the computing facilities.

\end{document}